\documentclass[a4paper,11pt]{article}
\usepackage{pos}

\newcommand{\cD}{\ensuremath{\mathcal D} }
\newcommand{\cDbar}{\ensuremath{\overline{\mathcal D}} }
\newcommand{\Ibb}{\ensuremath{\mathbb I} }
\newcommand{\cN}{\ensuremath{\mathcal N} }
\newcommand{\cQ}{\ensuremath{\mathcal Q} }
\newcommand{\cU}{\ensuremath{\mathcal U} }
\newcommand{\cUbar}{\ensuremath{\overline{\mathcal U}} }
\newcommand{\al}{\ensuremath{\alpha} }
\newcommand{\be}{\ensuremath{\beta} }

\newcommand{\vareps}{\ensuremath{\varepsilon} }
\newcommand{\ka}{\ensuremath{\kappa} }
\newcommand{\la}{\ensuremath{\lambda} }
\newcommand{\lalat}{\ensuremath{\la_{\text{lat}}} }
\newcommand{\muhat}{\ensuremath{\widehat\mu} }
\DeclareMathOperator{\Tr}{Tr}
\newcommand{\eq}[1]{Eq.~\ref{#1}}
\newcommand{\fig}[1]{Fig.~\ref{#1}}
\newcommand{\secref}[1]{Section~\ref{#1}}
\newcommand{\refcite}[1]{Ref.~\cite{#1}}

\title{Maximally supersymmetric Yang--Mills \\ \hfill in three dimensions}
\ShortTitle{Maximally supersymmetric Yang--Mills in three dimensions}

\author*{David Schaich}
\author{Angel Sherletov}
\affiliation{Department of Mathematical Sciences, University of Liverpool, Liverpool L69 7ZL, United Kingdom} 
\emailAdd{david.schaich@liverpool.ac.uk}

\FullConference{41st International Symposium on Lattice Field Theory (Lattice 2024) \\ 28 July -- 3 August 2024 \\ University of Liverpool}

\abstract{ 
  We present the latest results from our ongoing lattice field theory investigations of maximally supersymmetric Yang--Mills theory in three space-time dimensions, focusing on its non-perturbative phase diagram.
  Exploiting a lattice formulation that preserves a subset of the supersymmetry algebra at non-zero lattice spacing, we study the `spatial deconfinement' phase transition that holography relates to the transition between localized and homogeneous black branes in the dual quantum gravity.
  Fixing $N_L^2 \times 8$ lattice volumes and $N = 8$ colors in the SU($N$) gauge group, we consider four aspect ratios $\al = N_L / N_T \leq 4$ corresponding to $N_L = 16$, $20$, $24$ and $32$.
  The transition temperatures we determine are in good agreement with the low-temperature, large-$N$ holographic expectation $T_c \propto \al^3$.
}

\begin{document}
\maketitle

\section{\label{sec:intro}Introduction} 
Recent years have seen significant progress in numerical lattice field theory studies of supersymmetric gauge theories, exploiting formulations that exactly preserve a supersymmetry sub-algebra at non-zero lattice spacing.
See \refcite{Schaich:2022xgy} for a recent review.
These investigations are motivated by the many non-perturbative phenomena that play important roles in theoretical physics, not least the idea of `holographic' gauge/gravity duality that relates maximally supersymmetric Yang--Mills (SYM) to quantum gravity in higher-dimensional space-time.

Due to both the complexity of supersymmetric lattice field theories, and the need to consider the large-$N$ limit of the SU($N$) gauge group at strong 't~Hooft coupling $\la = N g_{\text{YM}}^2$, these investigations are computationally demanding.
Four-dimensional $\cN = 4$ SYM in particular poses additional challenges as an exactly conformal theory~\cite{Bergner:2021ffz, Schaich:2023war}.
As a result, recent efforts have focused on SYM theories in $p + 1$ dimensions with $p < 3$, still retaining the maximal number of $Q = 16$ supersymmetries.
Much of this work has dealt with the quantum-mechanical $p = 0$ case (most recently \refcite{Jha:2024rxz}, where many further references can be found), while Refs.~\cite{Giguere:2015cga, Kadoh:2017mcj, Catterall:2017lub, Jha:2017zad} have considered $p = 1$ during the past decade.

Here we instead turn our attention to three-dimensional SYM with $p = 2$, which offers a promising balance between computational costs and rich non-perturbative dynamics, including connections to quantum gravity via holographic dualities.
These explorations began with Refs.~\cite{Catterall:2020nmn, Sherletov:2022rnl}, which employed symmetric lattice volumes ($N_x \times N_y \times N_T$ with $N_x = N_y = N_T$) and confirmed that the bosonic action density behaves as expected from calculations of the free energy density in the homogeneous phase of the dual supergravity characterized by black D2 branes.
More recently~\cite{Sherletov:2023udh}, we have generalized our work to consider $N_L \times N_L \times N_T$ lattice volumes with different spatial and temporal extents, $N_L \neq N_T$.
This enables us to study the D2--D0 phase transition in the dual supergravity, which is expected to correspond to a `spatial deconfinement' transition at low temperatures in the large-$N$ limit.

In this proceedings we present first preliminary results for the critical temperature $T_c$ of this transition obtained using non-perturbative lattice field theory calculations.
In particular, we investigate $T_c$ for four different aspect ratios $\al = N_L / N_T \leq 4$, and observe good agreement with the holographic expectation $T_c \propto \al^3$~\cite{Morita:2014ypa, Catterall:2017lub}.
We begin in the next section by reviewing the lattice formulation we employ, and confirming that our numerical results pass several sanity checks (exhibiting thermal deconfinement and very small violations of a supersymmetric Ward identity).
Section~\ref{sec:results} presents our current results for the phase diagram, which remain preliminary in the sense that numerical calculations and analyses are still underway and may lead to changes prior to final publication.
However, the preliminary results appear reliable and robust, leading us to consider promising next steps in our concluding \secref{sec:conc}.

\section{\label{sec:lattice}Lattice formulation and numerical checks} 
Lattice regularization of SYM theories necessarily breaks the super-Poincar\'e algebra, which relates anti-commutators of supercharges to generators of infinitesimal translations that are not present in a discrete lattice space-time~\cite{Schaich:2022xgy}.
However, it is possible to identify a sub-algebra that can be preserved at non-zero lattice spacing, by changing variables to arrange the fields and supercharges into integer-spin representations of a `twisted rotation group' formed from the direct product of the ordinary rotations (in euclidean space-time) and the global R-symmetry of the theory.
This approach is known as `topological twisting', and is reviewed in detail by \refcite{Catterall:2009it}.
It requires $Q \geq 2^d$ supersymmetries in $d = p + 1$ dimensions, and provides at least one nilpotent twisted-scalar supercharge $\cQ$, preserving the sub-algebra $\{\cQ, \cQ\} = 0$ at non-zero lattice spacing.
This in turn significantly reduces the amount of fine-tuning required in lattice calculations~\cite{Schaich:2022xgy}.

Although we are interested here in maximal ($Q = 16$) SYM in three dimensions, it is convenient for us to follow Refs.~\cite{Catterall:2020nmn, Sherletov:2022rnl, Sherletov:2023udh} and proceed by employing a straightforward dimensional reduction of existing high-performance parallel software we have produced for four-dimensional $\cN = 4$ SYM~\cite{susy_code, Schaich:2014pda}.
That is, we formally carry out four-dimensional calculations, with lattice volumes $N_L \times N_L \times 1 \times N_T$.
The lattice action we use therefore features the same $\cQ$-exact and $\cQ$-closed terms as twisted continuum $\cN = 4$ SYM:
\begin{equation}
  \label{eq:action}
  \begin{split}
    S & = \frac{N}{4\lalat} \sum_n \Tr\left[\cQ \left(\chi_{ab}(n)\cD_a^{(+)}\cU_b(n) + \eta(n) \cDbar_a^{(-)}\cU_a(n) - \frac{1}{2}\eta(n) d(n) \right)\right] \\
      & \qquad -\frac{N}{16\lalat} \sum_n \Tr\left[\vareps_{abcde}\ \chi_{de}(n + \muhat_a + \muhat_b + \muhat_c) \cDbar_c^{(-)} \chi_{ab}(n)\right].
  \end{split}
\end{equation}
See \refcite{Catterall:2015ira} for full information about this expression.
In brief, the indices range from $1, \cdots, 5$, with the $Q = 16$ fermionic fields arranged as $\eta$, $\psi$ and $\chi_{ab} = -\chi_{ba}$ with 1, 5 and 10 components, respectively.
The finite-difference operators $\cD_a^{(+)}$ and $\cDbar_a^{(-)}$ involve the five-component complexified gauge links $\cU_a$ and $\cUbar_a$ into which both the gauge and scalar fields are arranged, which results in U($N$) rather than SU($N$) gauge invariance.

The 4d theory is formulated on the $A_4^*$ lattice with five symmetric basis vectors, which leaves us with three-dimensional SYM on the body-centered cubic ($A^*_3$) lattice after dimensionally reducing the $z$-direction.
In the temporal direction we impose thermal boundary conditions (BCs) --- periodic for the bosons and anti-periodic for the fermions --- while periodic BCs are used in all other directions.
This produces a skewed 3-torus~\cite{Catterall:2020nmn}, whose size can be expressed in terms of the dimensionless lengths
\begin{align}
  r_L & = L \la = N_L \lalat &
  r_T & = \be \la = N_T \lalat,
\end{align}
where $L$ and \be are dimensionful lengths, \la is the dimensionful 't~Hooft coupling in the continuum, and $\lalat = a\la$ is the dimensionless lattice 't~Hooft coupling (with lattice spacing `$a$').
Due to the skewed nature of the torus, the dimensionless temperature is~\cite{Catterall:2020nmn}
\begin{equation}
  \label{eq:T}
  T = \frac{4}{\sqrt{3}} \frac{1}{r_T}.
\end{equation}
In this setup, the continuum limit corresponds to extrapolating $N_L, N_T \to \infty$ while $\lalat \to 0$ so as to keep $r_L$ and $r_T$ fixed.

In order to carry out numerical calculations, we need to add to the lattice action~(\ref{eq:action}) two deformations, each of which softly breaks the twisted-scalar supersymmetry $\cQ$.
The first is a single-trace scalar potential that lifts the SU($N$) flat directions:
\begin{equation}
  \label{eq:single_trace}
  S_{\text{soft}} = \frac{N}{4\lalat} \mu^2 \sum_{n, a} \Tr\left[\bigg(\cUbar_a(n) \cU_a(n) - \Ibb_N\bigg)^2\right].
\end{equation}
The second deformation ensures that we obtain Kaluza--Klein dimensional reduction as opposed to Eguchi--Kawai volume reduction, by breaking the center symmetry in the reduced $z$-direction:
\begin{equation}
  \label{eq:center}
  S_{\text{center}} = \frac{N}{4\lalat} \ka^2 \sum_n \Tr\left[\bigg(\cU_z(n) - \Ibb_N\bigg)^{\dag}\bigg(\cU_z(n) - \Ibb_N\bigg)\right].
\end{equation}
In order to remove the soft supersymmetry breaking and recover the physical flat directions in the $\lalat \to 0$ continuum limit, we set the tunable parameters $\mu = \ka = \zeta\lalat$ with constant $\zeta$.
In this work we employ $\zeta = 0.5$, $0.6$ and $0.7$ for different aspect ratios $\al = r_L / r_T$ --- full details will be published in an open data release.

\begin{figure}
  \centering
  \includegraphics[width=0.7\linewidth]{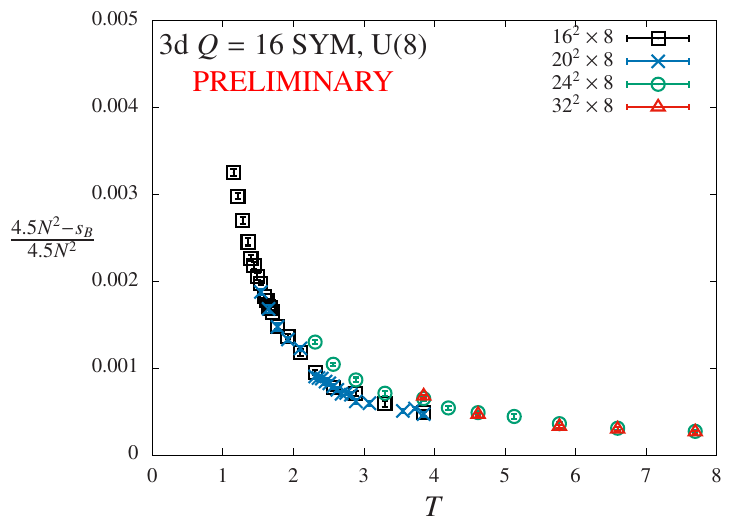}
  \caption{\label{fig:SB}Normalized violations of a $\cQ$-supersymmetry Ward identity involving the bosonic action, $s_B = 9N^2 / 2$, plotted vs.\ the dimensionless temperature from \eq{eq:T}.  The violations increase for the stronger 't~Hooft couplings at lower $T$, but remain under half a percent for all the calculations considered here.}
\end{figure}

Our numerical calculations use the standard rational hybrid Monte Carlo (RHMC) algorithm, which we have implemented in the publicly available parallel software package for lattice supersymmetry~\cite{susy_code} presented by \refcite{Schaich:2014pda}.
In the time since \refcite{Schaich:2014pda} appeared, we have implemented several improvements and extensions to this software, including an improved action~\cite{Catterall:2015ira}, $p = 0$ matrix models~\cite{Jha:2024rxz}, the deformation in \eq{eq:center}, and ongoing development of code for three-dimensional SYM with $Q = 8$ supercharges~\cite{Sherletov:2022rnl}.

There are two important checks that we discuss here before moving on to present our results for the phase diagram.
First, in all our calculations we quantitatively monitor the severity of the $\cQ$-supersymmetry breaking introduced above --- both from the two deformations as well as from the thermal BCs in the temporal direction.
We do this by computing the bosonic action $s_B$, which is fixed to a $\lalat$-independent value $s_B = 9N^2 / 2$ by a \cQ Ward identity.
Figure~\ref{fig:SB} plots normalized violations of this Ward identity vs.\ the dimensionless temperature $T$, observing that the violations are very small (well under half a percent) for all calculations considered here.
From this figure we can also conclude that Eqs.~\ref{eq:single_trace} and \ref{eq:center} are the dominant source of supersymmetry breaking, since the Ward identity violations become larger at lower temperatures where the 't~Hooft coupling is stronger, rather than at high temperatures where the thermal BCs have more effect.

\begin{figure}
  \centering
  \includegraphics[width=0.7\linewidth]{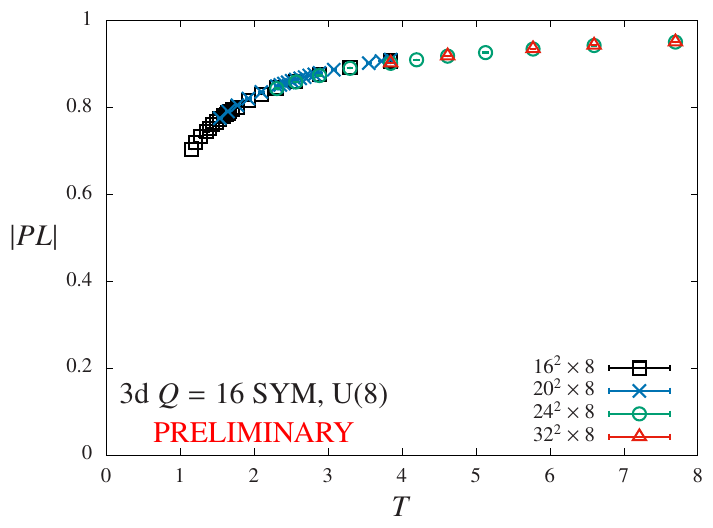}
  \caption{\label{fig:poly}The magnitude of the Polyakov loop vs.\ $T$.  For all calculations considered here, $|PL|$ is more than large enough to confirm the thermal deconfinement required for holographic duality to apply.}
\end{figure}

Second, in order for holography to relate this three-dimensional SYM field theory to black branes in higher-dimensional quantum gravity, the system must always be thermally deconfined.
We check this by monitoring the Polyakov loop (i.e., the Wilson line in the temporal direction), which is constructed from the unitary part $U_a$ of the complexified links $\cU_a$, extracted through a polar decomposition, $\cU_a = H_a \cdot U_a$~\cite{Catterall:2017lub, Catterall:2020nmn}.
We normalize the Polyakov loop so that its maximum possible value is unity.
In \fig{fig:poly} we see that the magnitude of the Polyakov loop is large ($|PL| > 0.7$) for all calculations considered here, confirming the thermal deconfinement needed for our results to have a holographic interpretation.

\section{\label{sec:results}Phase diagram results} 
While the systems we consider are always thermally deconfined, we expect a non-trivial `spatial deconfinement' phase structure.
At low temperatures (large $r_T$) and large $N$, holography predicts a first-order transition as $r_L$ decreases, from a large-$r_L$ phase of homogeneous black D2 branes to a small-$r_L$ phase of localized black holes (D0 branes)~\cite{Morita:2014ypa}.
On the field theory side, this is signalled by the Wilson lines in the two (symmetric) spatial directions, which we again construct from the unitary parts of the complexified gauge links.
While the precise $r_L$ at which the transition occurs is not known, there is a generic prediction for its temperature dependence~\cite{Morita:2014ypa},
\begin{equation}
  \label{eq:holo}
  T \propto r_L^{(p - 5) / 2} \qquad \implies \qquad T \propto \al^3,
\end{equation}
where we obtain the second expression by setting $p = 2$ and dividing both sides by $T^{3 / 2} \propto r_T^{-3 / 2}$.

\begin{figure}
  \centering
  \includegraphics[width=0.7\linewidth]{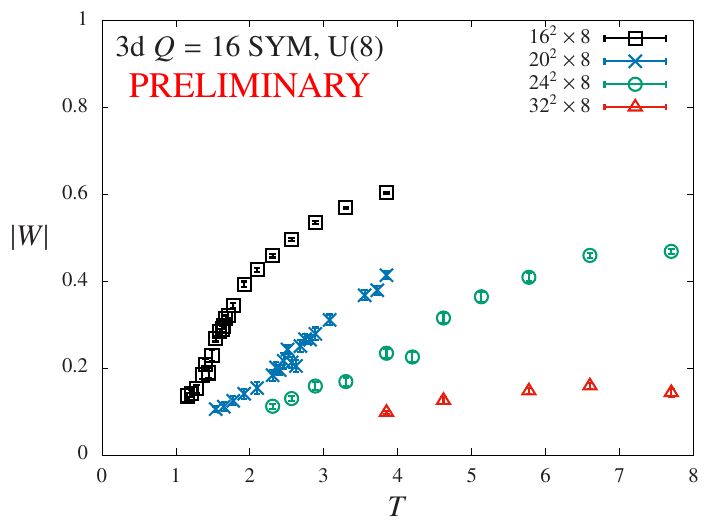} \\
  \includegraphics[width=0.7\linewidth]{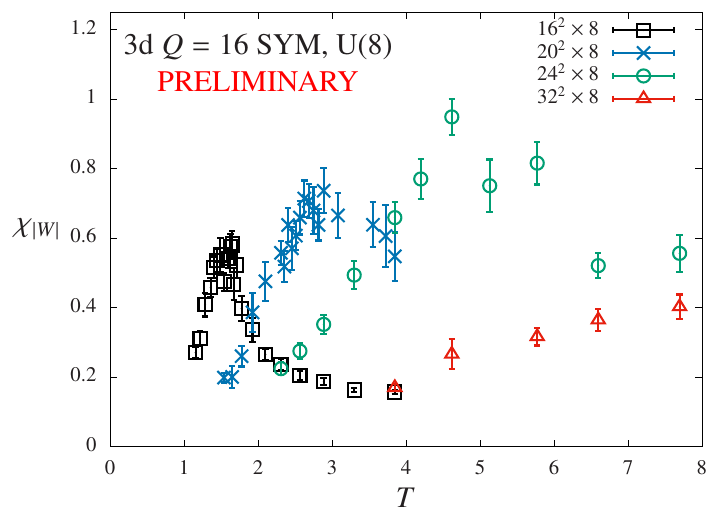}
  \caption{\label{fig:signals}Spatial Wilson line magnitudes (top) and the corresponding susceptibilities (bottom), both plotted vs.\ $T$.  Each point combines the Wilson lines in both the $x$- and $y$-directions.  The peaks in the susceptibilities for $\al = N_L / N_T \leq 3$ signal the spatial deconfinement transition.}
\end{figure}

\begin{figure}
  \centering
  \includegraphics[width=0.7\linewidth]{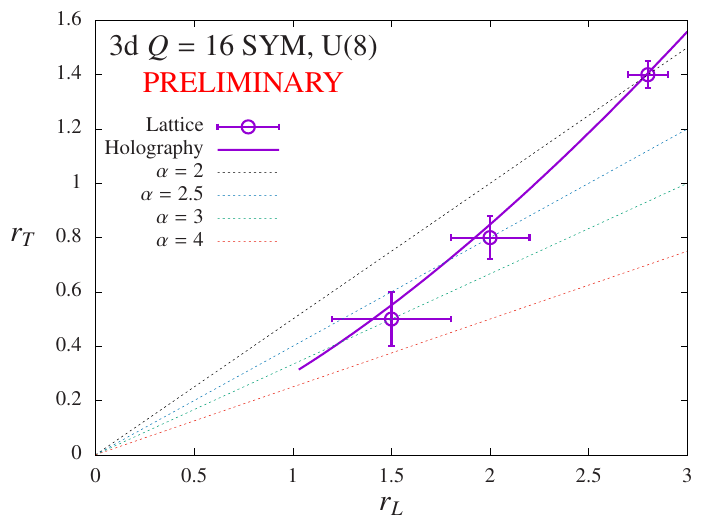}
  \caption{\label{fig:diag}Numerical lattice field theory results for the three-dimensional SYM phase diagram in the $r_T$--$r_L$ plane, compared to the holographic expectation \eq{eq:holo} in the form $r_T = c r_L^{3/2}$ (solid line).  We set $c = 0.3$ by hand (rather than fitting the points), which corresponds to $T = 0.21\al^3$.  The dotted diagonal lines show the trajectories with scan with fixed aspect ratio $\al = r_L / r_T$.}
\end{figure}

Our numerical approach is to fix the aspect ratio \al and simultaneously vary both $r_L$ and $r_T$, so that we scan along diagonal lines in the $r_T$--$r_L$ plane.
From \fig{fig:diag} we can see that larger aspect ratios should exhibit transitions at higher temperatures (smaller $r_T$ and $r_L$), due to \eq{eq:holo}.
Of course, as the temperature increases, that holographic prediction becomes less reliable, and will eventually break down.
At very high temperatures, we expect the fermions to pick up large thermal masses and decouple (completely removing supersymmetry), leaving us with a two-dimensional purely bosonic SU($N$) gauge--scalar system, about which a limited amount is known~\cite{Aharony:2005ew}.

In the calculations presented here, we consider $N_L^2 \times 8$ lattice volumes with fixed $N_T = 8$ and four different $N_L = 16$, $20$, $24$ and $32$ giving $\al = 2$, $2.5$, $3$ and $4$, respectively.
We also fix the number of colors to $N = 8$.
Our numerical calculations remain underway, so all the results presented here are preliminary and subject to change prior to final publication.

In \fig{fig:signals} we collect all our results for the Wilson line magnitudes (top) and the corresponding susceptibilities (bottom), combining the Wilson lines that wrap around the torus in either the $x$-direction or the $y$-direction.
For $\al \leq 3$, peaks are clearly visible in the susceptibilities, from which we can read off estimates for the transition temperatures: $T_c = 1.65(6)$ for $\al = 2$, $T_c = 2.9(4)$ for $\al = 2.5$, and $T_c = 4.6(9)$ for $\al = 3$.
These peaks get broader and noisier as \al increases, and especially for $\al = 3$ we are working to improve our determination of $T_c$ by accumulating more statistics and generating additional ensembles separated by smaller $\Delta r_T$.
In addition, we encounter greater difficulties with instabilities in the RHMC calculations as $T$ increases.
Although we increase $\zeta$ for larger \al (as discussed below \eq{eq:center}), we remain unable to access $T > 8$, which leaves the $\al = 4$ transition out of reach.

We collect our results for the transition temperatures in \fig{fig:diag}, now plotted in the $r_T$--$r_L$ plane rather than vs.\ $T = 4 / (r_T \sqrt{3})$.
We compare our three points to \eq{eq:holo} in the form $r_T = c r_L^{3/2}$, which corresponds to $T_c = 4c^2 \al^3 / \sqrt 3$. 
The solid line in \fig{fig:diag} exhibits good agreement with the numerical results.
To obtain it we have fixed $c = 0.3$ by hand (rather than fitting), so that $T_c \approx 0.21 \al^3$.
We can check the resulting $T_c \approx 1.66$ for $\al = 2$, $T_c \approx 3.2$ for $\al = 2.5$, and $T_c \approx 5.6$ for $\al = 3$ against the numbers in the previous paragraph.
This figure also suggests $r_T^{(c)} \lesssim 0.25 \longrightarrow T_c \gtrsim 16 / \sqrt{3} \approx 9.2$ for $\al = 4$, which is consistent with \fig{fig:signals} but might depart from the regime where low-temperature holography provides reliable predictions.

\section{\label{sec:conc}Conclusions and next steps} 
We have presented first preliminary results from numerical lattice field theory calculations for the critical temperature of the spatial deconfinement transition in three-dimensional SYM with maximal supersymmetry, observing good agreement with expectations from the holographic transition between homogeneous black D2 branes and localized black holes.
Our work, while still preliminary, is close to complete.
We continue accumulating more statistics to reduce noise in the higher-temperature transitions shown in Figs.~\ref{fig:signals} and \ref{fig:diag}, as well as adding more ensembles around the susceptibility peaks.
We are also analyzing the eigenvalue distributions of the spatial Wilson lines, which will confirm the homogeneous and localized natures of the phases on either side of the transition~\cite{Catterall:2020nmn}.

An interesting feature of \fig{fig:diag} is that uncertainties increase for higher temperatures, which is the opposite of the behavior observed in the two-dimensional case by Refs.~\cite{Catterall:2017lub, Jha:2017zad}.
That earlier work used a different lattice action, which may motivate experimentation with different lattice actions in the future.
Another well-motivated direction for future work is to adapt gradient flow techniques to three-dimensional SYM, which we expect will enable significant improvements in noise reduction at little computational cost~\cite{LatticeStrongDynamics:2020jwi}.

In the nearer term, there are several further calculations using our current setup that could improve our analyses of the phase diagram for three-dimensional SYM.
First, considering additional aspect ratios, such as $18^2\times 8$ or even $14^2\times 8$ lattice volumes, would strengthen the check of the holographic $T_c \propto \al^3$ dependence, by adding more points to \fig{fig:diag}.
Calculations with larger $N_T = 10$ and $12$ will enable extrapolations to the continuum limit, with computational costs that should remain within reach of the resources available to us.
Finally, checking whether the transitions we observe are in fact first order, as predicted by low-temperature holography, requires carrying out a scaling analysis in the number of colors $N$.
We have already investigated $N = 4$ and $6$, finding that for these gauge groups the instabilities discussed in the context of \fig{fig:signals} become unmanageable with our current lattice action.
While experimenting with different lattice actions may improve this situation, it would be more straightforward to carry out calculations with $N = 10$ and $12$.
However, because the computational costs grow faster than $N^3$ as $N$ increases~\cite{Catterall:2020nmn}, this approach would require significantly larger computing resources.

\vspace{24 pt} 
\noindent \textsc{Acknowledgments:}~We thank Navdeep Singh Dhindsa, Raghav Jha, Anosh Joseph and Toby Wiseman for recent and ongoing collaboration on three-dimensional $Q = 16$ lattice SYM.
Numerical calculations were carried out at the University of Liverpool and at the University of Cambridge through the STFC DiRAC facility.
DS was supported by UK Research and Innovation Future Leader Fellowship {MR/S015418/1} \& {MR/X015157/1} and STFC grants {ST/T000988/1} \& {ST/X000699/1}. \\[8 pt] 

\noindent \textbf{Data Availability Statement:} The data used in this work can be obtained by contacting DS.

\bibliographystyle{JHEP}
\bibliography{lattice24}
\end{document}